\documentclass{aa}             
\usepackage{amssymb,graphicx}  

\begin{document}

\thesaurus{06 (02.01.2 - 08.02.1 - 02.02.1 - 13.25.5 - 08.09.2 A0620-00)}

\title{Black hole X-ray binaries:\\
                A new view on soft-hard spectral transitions}
\author{F. Meyer \inst{1}, B.F. Liu \inst{1}\inst{,2}, E. Meyer-Hofmeister \inst{1}
}
\offprints{Emmi Meyer-Hofmeister}
\institute{Max-Planck-Institut f\"ur Astrophysik, Karl
Schwarzschildstr.~1, D-85740 Garching, Germany
\and
Yunnan Observatory, Academia Sinica. P.O.Box 110, Kunming 650011, China
} 

\date{Received:s / Accepted:}
\titlerunning {Black hole X-ray binaries:soft-hard spectral transitions}
\maketitle

\begin{abstract}

The theory of coronal evaporation predicts the formation of an
inner hole in the cool thin accretion disk for mass accretion rates
below a certain value ($\approx$ 1/50 of the Eddington mass
accretion rate) and the sudden disappearance of this hole when
the mass accretion rate rises above that value.
The inner edge of the standard thin disk then suddenly shifts inward
vvvvfrom about a few hundred Schwarzschild radii to the last stable orbit.
This appears to quantitatively account for the observed transitions
between  hard and soft spectral states at critical luminosities.
Due to the evaporation process the matter accreting in the geometrically
thin disk changes to a hot coronal flow which proceeds towards the
black hole as an advection-dominated accretion flow (ADAF; for a
review see Narayan et al. 1998). 
\end{abstract}

\section{Introduction}
For a decade it has been known that the spectra of X-ray novae
show changes from a soft state at high luminosity to a hard 
state when the luminosity has declined during the outburst
(Tanaka 1989). The persistent canonical black hole
system Cyg X-1 also undergoes occasional transitions between its standard low
luminosity (hard) state and a soft state (see Fig. 1). Such changes
between the two spectral states have been observed for several
systems, regardless of whether the compact object is a neutron star
(Aql X-1, 1608-522) or a black hole (GS/GRS 1124-684, GX 339-4)
(Tanaka \& Shibazaki 1996). Here we concentrate on black hole sources.  
Observations show that the phenomenon always occurs at a luminosity around
$10^{37}\rm{erg/s}$, which corresponds to a mass accretion rate of
about $10^{17}\rm{g/s}$ (Tanaka 1999).

The two spectral states are thought to be related to different states
of accretion:
(1) the soft spectrum originates from a thin disk which extends down
to the last stable orbit plus a corona above the disk, 
(2) the hard spectrum originates from a thin disk outside a
transition radius $r_{tr}$ and a coronal flow/ ADAF
inside. The spectral
transitions of Nova Muscae 1991 and Cygnus X-1 were modelled based on
this picture by Esin et al. (1997, 1998). The value of $r_{tr}$ was
taken as the maximal distance $r$ for which an ADAF with that accretion
rate can exist (``strong ADAF proposal", Narayan \& Yi 1995). 
We determine the location of the inner edge of the thin disk from the
equilibrium between it and the corona above. 

\section{Generation of the coronal flow}
\subsection{Evaporation}
The equilibrium between the cool accretion disk and the corona
above (Meyer \& Meyer-Hofmeister 1994) is established in the following
way. Frictional heat released in the
corona flows down into cooler and denser transition layers. There it
is radiated away if the density is sufficiently high. If the density is
too low, cool matter is heated up and evaporated into the corona until
an equilibrium density is established (Meyer 1999).

Mass drained from the corona by an inward drift is replaced by mass
evaporating from the thin disk as the system establishes a stationary state.
When the evaporation rate exceeds the mass flow rate in the cool disk
the disk terminates. Inside only a hot coronal flow exists.

\begin{figure}[ht]
\includegraphics[width=8.3cm]{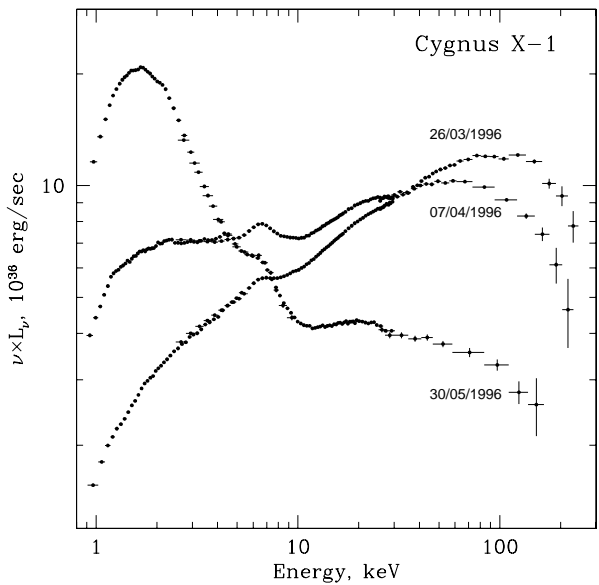}
\caption{Transition from the hard spectrum on 26/3/1996 to a soft
spectrum on 30/5/1996, observed for Cygnus X-1 (from M. Gilfanov,
E. Churazov, M.G. Revnivtsev, in preparation)}
\end{figure}

\subsection{Physics of the corona}
Mass flow in the corona is similar to that in the thin disk.
Differential (Kepler-like) rotation causes transfer of angular momentum
outwards and mass flow inwards. The corona is geometrically much
thicker than the disk underneath. Therefore sidewise energy
transport is not negligible. Sidewise advection, heat conduction
downward, radiation from the hot optically thin gas flow and 
wind loss are all important for the equilibrium between corona and thin
disk. A detailed  description would demand the solution of a set of partial
differential equations in radial distance $r$ and vertical height $z$.
In particular a sonic transition requires treatment of a free
boundary condition on an extended surface.

From simplified modelling and analysis we find the
following pattern of coronal flow. When a hole in the thin disk
exists there are three regimes with increasing distance from the black hole.
(1) Near the inner edge of the thin disk
the gas flows towards the black hole. (2) At
larger $r$ wind loss is important taking away about 20\% of the total
matter inflow. (3) At even larger distances some matter
flows outward in the corona as a consequence of conservation of angular
momentum. One might compare this with the
flow in a ``free'' thin disk without the tidal forces acting in a
binary. In such a disk matter flows inward in the inner region and
outward in the outer region, with conservation of the total mass and
angular momentum (Pringle 1981).

\subsection{Model}
We model the equilibrium between corona and thin disk in a simplified
way. This is possible since the evaporation process is concentrated
near the inner edge of the thin disk. Thus the corona above the
innermost zone of the disk dominates the global structure. Further
inward there is no thin disk anymore. The representative dominant region
from $r$ to $r$+$\Delta r$ has to be chosen such that evaporation
further outward is not important. One incorporates the effects of
frictional heat generation, conduction, radiation, sidewise loss of
energy and wind loss at large height into this one zone
( ``one-zone-model''). A set of ordinary differential equations for mass,
motion, and energy with boundary conditions at the bottom (downward
thermal flux \,-\, pressure relation) and at the top (sonic
transition) {\it{uniquely}}
determine mass accretion rate, wind loss and temperature in the corona
as a function of radius. We restrict the analysis to a stationary corona.

The evaporation process was first investigated for disks in dwarf nova
systems (Meyer \& Meyer-Hofmeister 1994, Liu et al 1995). The
situation is similar for disks around black holes (Meyer 1999). The
coronal gas flowing into the hole and replaced by evaporation from the
disk is understood as the supply for an ADAF which was used successfully to
model the spectra of several black hole sources. A recent review by
Narayan et al. (1998) gives a detailed description of accretion in the
vicinity of a black hole.

\section{Computational results}

\subsection{The critical mass flow rate ${\dot M_{\rm{crit}}}$} 
We use the same equations as Liu et al. (1995).
The efficiency of evaporation at given distance $r$ from the
compact star determines the location of the inner edge of the thin
disk $r_{\rm{tr}}$. The relation between the mass flow rate ${\dot M}$
in the disk and $r_{\rm{tr}}$ was now computed also for black
hole systems. In Fig. 2 we show this relation for a 6 $M_\odot$
black hole (viscosity parameter $\alpha$=0.3).

Up to now only the decreasing branch was known and investigated.
The interesting new feature is that the efficiency
of evaporation reaches a maximum. This means that as the mass
accretion rate in the disk is increased the inner edge moves inward, but
if the rate exceeds a critical value $\dot M_{\rm{crit}}$ the
thin disk can no longer be fully depleted by evaporation (for this accretion rate the
inner disk edge is at about 340 Schwarzschild radii). The thin disk then
extends inward towards the last stable orbit.

\begin{figure}[ht]
\includegraphics[width=7.5cm]{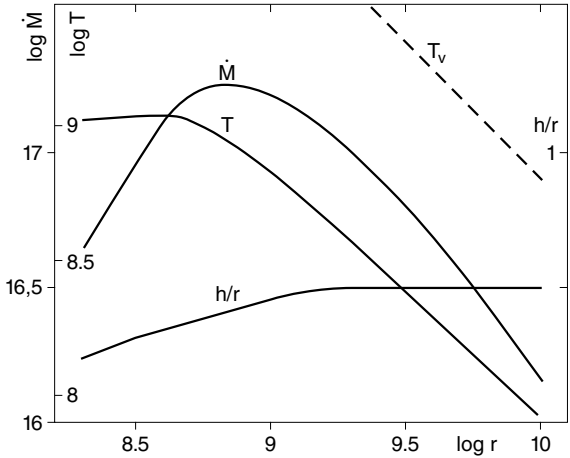}
\caption{
Solid lines: rate of inward mass flow $\dot M$(in g/s) in the
corona (= evaporation rate), maximum temperature in the corona and $h/r$
(h pressure scaleheight) at the inner edge $r$=$r_{\rm{tr}}$ of the
standard thin disk. Dashed line: virial temperature.}
\end{figure}

The temperature in the corona increases with decreasing radius, but
reaches a saturation value where the coronal mass flow reaches maximum.
The value h in Fig. 2 is the height where the pressure has
decreased by 1/e.

\subsection{What causes the maximum of the coronal mass flow rate?}
A change in the physical process that removes the heat released by friction
is the cause for the maximum of the coronal mass flow rate seen in
Fig. 2. A dimensional analysis of the equations
yields the following result. 
For large inner radii coronal heating is balanced by inward advection 
and wind loss. This fixes the coronal temperature at about 1/8 of the
virial temperature $T_{\rm{v}}$
($\Re T_{\rm v}/\mu=GM/r$, $\Re$ gas constant, $\mu$ molecular weight,
$G$ gravitational constant) (see Fig. 2). Downward heat conduction and
subsequent radiation 
in the denser lower region play a minor role for the energy loss though 
they always establish the equilibrium  density in the corona above the 
disk.

With rising temperature, thermal heat conduction removes an increasing
part of the energy released and finally becomes dominant.
For optically thin bremsstrahlung the temperature saturates at a
universal value defined by a combination of conductivity and
radiation coefficients, the Boltzmann and the gas constant, and the
non-dimensional $\alpha$-parameter of friction, (see Fig. 2).  Dimensional
analysis of the equations yields the rate of mass accretion through the 
corona as a function of temperature divided by the
Kepler frequency $(GM/r^3)^{1/2}$. For small radii this gives the 
$r^{3/2}$ law in Fig. 2.

The maximum accretion rate occurs where the 
sub-virial temperature for large radii reaches the saturation temperature 
for small radii. Since the virial temperature is proportional to $M/r$, 
this radius $r_{\rm{crit}}$ is proportional to $M$. Then the
accretion rate, proportional to the inverse of the Kepler frequency, also 
becomes proportional to $M$.

\subsection{Approximations used for our model} 
Synchrotron and Compton cooling have been left out. Synchrotron cooling
is non-dominant as long as the magnetic energy density stays below roughly
1/3 of the pressure. Compton cooling and heating by photons from the
disk surface and from the accretion centre are non-dominant at all
distances larger than that of the peak of the coronal mass flow rate,
$r\ge r_{\rm{crit}}$ ($\approx  340 {R_s}$, Fig. 2). They become
important for smaller radii.

The conductive flux remains small compared to the upper limit,
the transport by free streaming electrons, so that classical
thermal heat conduction is a good approximation.
We have neglected lateral heat inflow by thermal conduction. 
This term is small compared to the dominant advective and wind
losses at large radii, and vanishes
when the temperature becomes constant at small radii. 

Temperature equilibrium between electrons and ions requires that
the collision times between them remains shorter than 
the heating timescale. This holds for $r\ge r_{\rm{crit}}$, but
the condition fails for $r < r_{\rm{crit}}$ where a two temperature
corona can develop.

Tangled coronal magnetic fields
could reduce electron thermal conductivity. We note
however that reconnection and evaporation tend to establish a rather
direct magnetic path between disk and corona.

\section{Spectral transitions}

\subsection{Predictions from the evaporation model}
At maximum luminosity of an X-ray nova outburst the mass accretion
rate is high and the thin disk extends inward to the last stable
orbit. A corona exists above the thin disk, but the mass flow in the
thin disk is so high that no hole appears.
In decline from outburst the mass accretion rate decreases. When
${\dot M_{\rm{crit}}}$ is reached a hole forms at $\approx$
340 ${R_s}$ and the transition soft/hard occurs.
If the mass accretion rate varies up and down as in high-mass X-ray
binaries we expect hard/soft and soft/hard transitions. 
In Fig. 3 we show the expected behaviour schematically.

The descending branch for smaller $r$ indicates the possibility that an
interior disk could form. We note that a gap exists between the exterior
standard thin disk and an interior disk. In this gap the flow assumes
the character of an ADAF with different temperature of ions and
electrons, due to its high temperature and poor collisional
coupling. This provides the possibility that the interior disk fed by
this flow has a two temperature corona on top, different from a
standard thin disk plus corona in the high state. We will discuss this
in a further investigation.

\begin{figure}[ht]
\includegraphics[width=8.3cm]{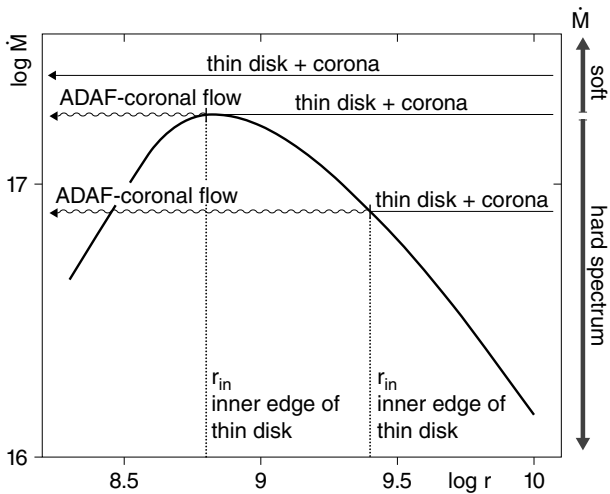}
\caption{
Evaporation rate ${\dot M(r)}$ as in Fig. 2. Inward extension of the standard
thin disk for 3 different mass flow rates ${\dot M}$ in the thin
disk (schematic). Note
that the standard thin disk reaches inward towards the black hole if 
${\dot M} \ge {\dot M_{\rm{crit}}}$. Shown also the type of spectrum,
soft or hard, related to ${\dot M}$}.
\end{figure}

\subsection{Comparison with observations}

The three persistent (high-mass) black hole  X-ray sources LMC X-1,
LMC X-3 and Cyg X-1 show a different behaviour. LMC X-1 is always  
in the soft state (Schmidtke et al. 1999). For LMC X-3, most of the time in the
soft state, recently recurrent hard states have been detected 
(Wilms et al. 1999). Cyg X-1 spends most of its time in the hard
state with occasional transitions to the soft state (see e.g. Fig. 1).
This can be interpreted as caused by different long-term mean mass transfer
rates: the highest rate (scaled to Eddington luminosity) in LMC X-1,
the lowest in Cyg X-1, and in between in LMC X-3.
Transient sources show a soft/hard transition during the decay from
outburst. The best studied source is the X-ray Nova Muscae 1991
(Cui et al. 1997).

The transition always occurs around $L_X\approx10^{37}$ erg/s (Tanaka 1999).
Our value for the critical mass accretion rate for a 6 $M_\odot$ black
hole, $10^{17.2}$g/s, corresponds to a standard accretion disk
luminosity of about $10^{37.2}$ erg/s. This is very close agreement.

For accretion rates below ${\dot M_{\rm{crit}}}$ the location of the
inner edge of the standard thin disk derived from the evaporation
model also agrees with observations (Liu et al. 1999).

At the moment of spectral transition our model predicts the inner edge
near 340 Schwarzschild radii. The observed timescale for the spectral
transition of a few days (Zhang et al. 1997) agrees with the time one
obtains for the formation of a disk at 340 ${R_s}$ with an accretion
rate ${\dot M_{\rm{crit}}}$.

But even in the low state X-ray observations of a reflecting component
indicates the existence of a disk further inward, at 10 to 25 ${R_s}$
(Gilfanov et al. (1998), Zycki et al. (1999)). This might point to
a non-standard interior disk as discussed above and explain why the
spectral transitions in Cygnus X-1 could be well fitted by  Esin et
al. (1998) with a disk reaching inward to $\le 100 {R_s}$.

\section{Conclusions}

We understand the spectral transition as related to a critical mass
accretion rate. For rates ${\dot M} \ge {\dot M_{\rm{crit}}}$ 
(the peak coronal mass flow rate) the standard disk reaches inward
to the last stable orbit and the spectrum is soft. Otherwise the ADAF
in the inner accretion region provides a hard spectrum.
At ${\dot M_{\rm{crit}}}$ the transition between dominant advective
losses further out and
dominant radiative losses further in occurs. Except for the difference
between the sub-virial temperature of the corona and the
closer-to-virial temperature of an ADAF of the
same mass flow rate, this same critical radius is predicted by the
``strong ADAF proposal"
(Narayan \& Yi (1995). In general however, the strong ADAF
proposal results in an ADAF region larger than that which the evaporation
model yields.

The transition between the two spectral states  has been observed
for black hole and neutron star systems, in persistent and transient
sources (Tanaka \& Shibazaki 1996, Campana et al.1998). This
points to similar physical accretion processes. Menou et al. (1999) already
discussed the accretion via an ADAF in neutron star transient
sources. Our results should also be applicable.

The relations for a 6 $M_\odot$ black hole plotted in Fig. 2 can be
scaled to other masses: in units of
Schwarzschild radii and Eddington accretion rates the plot is
universal. The application to disks around supermassive black holes
implies interesting conclusions for AGN.

\begin{acknowledgements}
We thank Marat Gilfanov, Eugene Churazov and Michael Revnivtsev for the
spectral data of Cygnus X-1.
\end{acknowledgements}


\begin{thebibliography}
{}
\bibitem{ref:5} Campana S., Colpi M., Mereghetti S., 1998, A\&A
Rev. 8, 269
\bibitem{ref:10} Cui W., Zhang S.N., Focke W. et al., 1997, ApJ 484,383
\bibitem{ref:19} Esin A.A., McClintock J.E, Narayan R., 1997, ApJ 489,
865
\bibitem{ref:30} Esin A.A., Narayan R., Cui W. et al., 1998, ApJ
505, 854
\bibitem{ref:33} Gilfanov M., Churazov E., Sunyaev R., 1998, in: 18th
Texas Symposium on Relativistic Astrophysics and Cosmology, eds.
A.V. Olinto et al.; World Scientific, p.735
\bibitem{ref:35} Liu B.F., Yuan W., Meyer F. et al., 1999, ApJ 527, L17
\bibitem{ref:40} Liu F.K., Meyer F., Meyer-Hofmeister E., 1995, A\&A
300, 823
\bibitem{ref:50} Menou K., Esin A., Narayan R. et al., 1999, ApJ 
520, 276
\bibitem{ref:52} Meyer F., 1999, in: Proc. of Disk Instabilities,
eds. S. Mineshige and J.C. Wheeler, Univ. Academic Press, Kyoto, p.209  
\bibitem{ref:55} Meyer F., Meyer-Hofmeister E. 1994, A\&A 288, 175
\bibitem{ref:81} Narayan R., Mahadevan R., Quartaert E., 1998, in:
The Theory of Black Hole Accretion Discs, eds. M.A. Abramowicz et al.,
Cambridge University Press, p.148
\bibitem{ref:82} Narayan  R., Yi, 1995,  ApJ 452, 710
\bibitem{ref:83} Pringle J.,1981, Ann. Rev. Astron. Astrophys. 137
\bibitem{ref:831} Schmidtke P.C., Ponder A.L., Cowley A.P., 1999,
Astron.J., 117, 1292
\bibitem{ref:84} Tanaka Y.,1989, in: Proc. 23rd ESLAB Symp. Two-Topics
X-Ray Astronomy, Bologna ESA SP-296, p.3
\bibitem{ref:85} Tanaka Y.,1999, in: Proc. of Disk
Instablities in Close Binary Systems, eds. S. Mineshige and
J.C. Wheeler, Universal Academic Press, Kyoto, p.21
\bibitem{ref:87} Tanaka Y., Shibazaki N., 1996, ARA\&A, 34, 607
\bibitem{ref:90} Wilms J., Nowak M.A., Pottschmidt K., et al. 1999,
astro-ph 9910508
\bibitem{ref:95} Zhang S.N., Cui W., Harmon B.A. et al., 1997, ApJ
477, L95
\bibitem{ref:100} Zycki T., Done C., Smith D.A., 1999, MNRAS 305, 231

\end{thebibliography}
\end{document}